\documentclass[preprint,12pt,epsfig]{revtex4}

\usepackage {graphicx}

\usepackage {amsmath}

\usepackage {amsfonts}

\usepackage {amssymb}

\usepackage {mathrsfs}

\usepackage{color}

\newcommand {\be}{\begin{eqnarray}}

\newcommand {\ee}{\end{eqnarray}}

\begin{document}


\title {Euclidian crystals in many-body systems: breakdown of Goldstone's theorem}

\author {S.I. Mukhin}
\email{sergeimoscow@online.ru;tel.+79031898657}
\affiliation {Theoretical Physics and Quantum Technologies Department, Moscow Institute for Steel and Alloys, Leninskii avenue 4,
119049 Moscow, Russia}
\author{{\it{Running head}}: Gapped Goldstone modes}



\begin{abstract}
It is proven via modification of the Nambu-Goldstone theorem, based on the Ward-Takahashi identity, that Goldstone's bosons of the quantum order could be massive. The quantum order parameter (QOP), introduced previously as  the instantonic crystal in Euclidian space (Mukhin 2011), breaks Matsubara's time translational invariance while possessing zero classical expectation value and vanishing scattering cross-section ("hidden order"). The amplitude of the mass-gap can be calculated from the eigenmodes spectrum of the effective Euclidian action for a particular self-consistent QOP solution in the fermionic repulsive Hubbard model. Theoretical results are discussed in relation with possible linking together of the two phenomena that previously were considered separately, i. e. "hidden order" fingerprints  in ARPES measurements and "neutron resonance" feature in the magnetic spectrum in lightly hole-doped copper oxides.\\ 

\end{abstract}

\maketitle

\section*{1. Introduction}
\label{sec:Introduction}

In accord with Nambu-Goldstone theorem  \cite{amit} spontaneous breaking of a global continuous symmetry of the system by a mean-field order parameter causes appearance of massless Goldstone bosons, called also gapless Goldstone modes.  It is demonstrated below that the theorem can be violated, i.e. the Goldstone modes possess finite gap, in the case of  a novel Euclidian crystal ("quantum ordered") state of interacting fermions proposed recently \cite{mukhin}. This result may lead to important consequences when considered in combination with another peculiar properties of the quantum order parameter  (QOP) found recently \cite{mukhin}:   QOP's  Green's function is finite and periodic along the Matsubara's axis, but Wick-rotated to the axis of real frequencies, possesses infinite periodic "chain" of \textit{second order poles} along this axis. This property of QOP leads to a remarkable physical consequence: {\it{QOP does not scatter anything in Minkowski world}}! Hence,  the present theory, we believe, may naturally indicate that two effects:  "hidden order" revealed in ARPES spectra of fermi-excitations  \cite{pg1,pg2} and "neutron resonance" feature 
\cite{hg,hg1,hg2,hg3} at finite frequency in the magnetic (bosonic) excitations spectrum in lightly hole-doped copper oxides, could be emerging fingerprints of the QOP state.
In Section 2 the derivation is presented. In Section 3 the peculiar manifestation of the QOP self-consistent solution is described for an "electromagnetic Euclidian crystal" predicted 
recently \cite{mf} in the array of Josephson junctions coupled via electromagnetic field in the superconducting resonator in the form of non-classical phonon states . In the Conclusions  we summarize the open problems and possible ways to solve them.

\section*{2. Generalized Goldstone theorem for QOP: finite frequency resonance?}
\label{sec:Goldstone}

It is useful first to describe shortly previously introduced concept of QOP \cite{mukhin,mukhin1}. For this purpose we use weakly-coupled repulsive-$U$ Hubbard model of electronic system:
\be
H=-\sum_{\langle i,j\rangle,s}\left(t_{i,j}{c_{i,s}}^{+}c_{j,s}+H.c.\right)+U\sum_{i}\hat{n}_{i,\uparrow}\hat{n}_{i,\downarrow}.
\label{H}
\ee
\noindent Here $t_{i,j}$ is electron hopping integral, $\langle i,j\rangle$ denote lattice sites, and ${c_{i,s}}^{+},\, c_{j,s}$ and $n_{i,s}$ are fermionic creation, annihilation and density operators respectively, and $s$ is fermionic spin variable ($s=\pm 1$ correspond to $\uparrow$ and $\downarrow$ spin-states). To study the QOP formation we follow general strategy of the phase transition theory and rewrite the Hubbard Hamiltonian $H$ from Eq. (\ref{H}) in a decoupled single-particle form using Hubbard-Stratanovich (HS) decoupling field\cite{hertz}:

\be
H_M=-\sum_{\langle i,j\rangle,s}\left(t_{i,j}{c_{i,s}}^{+}c_{j,s}+H.c.\right)+\sum_{i,s}sM(\tau,i)\hat{n}_{i,s}
\label{HS}
\ee
\noindent Here a SDW-type HS-field is chosen for definiteness. The Euclidian action $S$, and partition function $Z$ of electrons are expressed via (path)-integrals:
\be
\label{Z}
&Z=\displaystyle\int{\cal{D}}\bar{\Psi}{\cal{D}}\Psi{\cal{D}}{M(\tau,\bf{r})} exp{\left\{-S\right\}}\label{euclid}\\
&S=\displaystyle\int\int_0^{\beta}{d^{D}{\bf{r}}d\tau  \left[\bar{\Psi}\partial_\tau\Psi+H_{M}+\dfrac{|M(\tau,\bf{r})|^2}{4U}\right]}
\label{S}
\ee
\noindent  where $\beta=1/T$. In Eqs. (\ref{Z}), (\ref{S}) summation over the D-dimensional crystal-lattice sites $i$ is substituted with continuum integration $\int d^D{\bf{r}}$, and the Plank's constant is the unit of action ($\hbar=1$). In Eq.(\ref{Z}) the path-integration  substitutes the trace over diagonal elements of the Hamiltonian matrix, and hence it is performed over $\tau$-periodic HS-fields:
\begin{eqnarray}
{M}\left(\tau+{1}/{T},{\bf{r}}\right)={M}(\tau,{\bf{r}}),
\label{SHM}
\end{eqnarray}\noindent Formally, integration $\int {\cal{D}}{M(\tau,\bf{r})}$ in Eq. (\ref{Z}) turns $H_M$ from Eq. (\ref{HS}) back into $H$ from Eq. (\ref{H}). In case of a classical phase transition the overwhelming contribution to partition integral $Z$ comes from the $\tau$-independent HS field, i.e.{{\it{ the minimum of the Euclidian action $S$ is achieved with some particular $\tau$-independent function $M(\bf{r})$, that constitutes well known classical (mean-field) order parameter}}, COP. The minimum of $S$ condition, from which the COP is found, is called self-consistent mean-field equation and was first introduced by P. Weiss \cite{weiss} for ferromagnetic domains. As it was shown \cite{mukhin}, besides COP {\it{the Hubbard-Stratanovich field that minimizes the Euclidian action $S$ may become a QOP, being $\tau$-periodic function with zero mean}}:
\be
\langle M(\tau,{\bf{r}})\rangle_{1/T}=0.
\label{me}
\ee
 \noindent 
A proper theoretical description of the ordered state with QOP is based on the reducing the path-integration over Grassmanian fields $\bar{\Psi}\,,\Psi$ in the partition function $Z$ to solution of the Dirac-type equation with HS field playing the role of $\tau$-periodic potential $ M(\tau,\bf{r})$. Then, HS field $ M(\tau,\bf{r})$ is found that minimizes the Euclidian action $S$ and manifests QOP. This plan is accomplished in two steps \cite{mukhin}. First, a spectrum $\{\alpha_n\}$ of the quasi-energies (Floquet indices) of the  time-dependent Hamiltonian Eq. (\ref{HS}) is found \cite{neveu, mukhin}. The Floquet indices label solutions of the corresponding Dirac equation with $\tau$-dependent coefficients $ M(\tau,\bf{r})$ from Eq. (\ref{SHM}) entering Hamiltonian $H_M$:
\begin{eqnarray}
&(\partial_{\tau}+\hat{H}_{M})\vec{\psi}_n=0\,, 
\label{bloch}
&\vec{{\psi}}_n(\tau+{1}/{T})=e^{-\alpha_n}\vec{\psi}_n(\tau).
\label{FQF}
\end{eqnarray}
\noindent Second, provided the indices $\alpha_n$ are known, the partition function in Eq. (\ref{Z}) can be expressed as\cite{neveu,mukhin}:
\be
Z=\int{\cal{D}}{M(\tau,\bf{r})} exp{\{-S_{F}\}}\prod_n \cosh\left(\dfrac{\alpha_n}{2}\right)
\label{ZM}
\ee
\noindent where $S_{F}$ is the action of bosonic field $M(\tau,\vec{r})$:

\be
S_F=\int\int_0^{\beta}{d^{D}{\bf{r}}d\tau  \left[\dfrac{|M(\tau,\bf{r})|^2}{4U}\right]}
\label{SF}
\ee
\noindent The QOP is periodic function $M^{QOP}(\tau,{\bf{r}})$ that minimizes the total action: 
\be
&&\delta_{M(\tau,\vec{r})}{\left\{ S_F-\sum_n ln\left\{\cosh\left(\dfrac{\alpha_n}{2}\right)\right\}\right\}}=0,
\label{self}
\ee
\noindent  and obeys Eqs. (\ref{SHM}), (\ref{me}).   Here we consider weak-coupling limit ($U\ll t_{i,j}$) of Eq. (\ref{HS}) assuming that $\vec{r}$-dependence of $M(\tau,\vec{r})$ is harmonic:
\begin{eqnarray}
{M}(\tau,{\bf{r}})={M}(\tau)e^{i\vec{Q}\vec{r}}+{M}^{*}(\tau)e^{-i\vec{Q}\vec{r}},
\label{SDW}
\end{eqnarray}

\noindent where $\vec{Q}$ is a single SDW wave-vector. Simultaneously, in order to apply the path-integral approach \cite{fradkin}, we introduce a source field 
$h(\tau,{\bf{r}})$ corresponding to the QOP ${M}(\tau,{\bf{r}})$:

\begin{eqnarray}
{h}(\tau,{\bf{r}})={h}(\tau)e^{i\vec{Q}\vec{r}}+{h}^{*}(\tau)e^{-i\vec{Q}\vec{r}},
\label{SDW}
\end{eqnarray}

\noindent Now, following derivation in \cite{fradkin},  but with imaginary time dependent source field $\vec{h}(\tau,{\bf{r}})$ , we add the symmetry breaking source term to the Hamiltonian:
 \begin{eqnarray}
H_M=\sum_{q,s}E_q{c^{+}_{q,s}}c_{q,s}+\left(c^{+}_{q+Q,s}[M(\tau)+h(\tau)]s c_{q,s}+H.c.\right)
\label{HF}
\end{eqnarray}
\noindent
Now shifting the functional integration variable in (\ref{ZM}) $M\rightarrow M'$:
\begin{eqnarray}
{M'}(\tau,{\bf{r}})=\left({M}(\tau)+h(\tau)\right)e^{i\vec{Q}\vec{r}}+\left({M}^{*}(\tau)+h^{*}(\tau)\right)e^{-i\vec{Q}\vec{r}},
\label{SDW}
\end{eqnarray}
\noindent and making formal summation in (\ref{ZM}) over the Floquet indices under "time"-dependent field ${M'}(\tau,{\bf{r}})$ leading to $S_{eff}(M'(\tau,\bf{r}))$, we find the effective action of the QOP and the generating functional $F(h)$:
\be
&&Z=exp{\left[-F(h)-\int{d\tau\int d\bf{r}\dfrac{|h(\tau,\bf{r})|^2}{4U}}\right]}=\int{\cal{D}}{M'(\tau,\bf{r})}exp{\left[-S_{eff}(M'(\tau,\bf{r})) \right.} \nonumber\\
&&{\left. +\int d\tau\int d\bf{r}\dfrac{|M'(\tau,\bf{r})h(\tau,\bf{r})|}{2U}-\int d\tau \int d\bf{r}\dfrac{|h(\tau,\bf{r})|^2}{4U} \right]}
\label{ZME}
\ee
\noindent The wellknown property of the generating functional relates its variational derivative with the average value of the order parameter $\bar{M}(\tau,\bf{r})$(not necessarily spontaneous)  :
\be
\dfrac{\delta F(h)}{\delta h(\tau,\bf{r})}=\bar{M}(\tau,\bf{r})
\label{dfh}
\ee
\noindent By the Legendre transform one obtains free energy of the system $\Gamma(\bar{M})$:
\be
\Gamma(\bar{M})+F(h)=\int{d\tau\int d\bf{r}\bar{M}(\tau,\bf{r})h(\tau,\bf{r})}
\label{G}
\ee
\noindent together with the relation \cite{amit}:
\be
\dfrac{\delta \Gamma(\bar{M})}{\delta \bar{M}(\tau,\bf{r})}=h(\tau,\bf{r})
\label{dgm}
\ee
\noindent Then the spontaneous order parameter obeys the following equations:
\be
\dfrac{\delta \Gamma(\bar{M})}{\delta \bar{M}(\tau,\bf{r})}\displaystyle |_{\bar{M}=v(\tau,\bf{r})}=0;\;\dfrac{\delta F(h)}{\delta h(\tau,\bf{r})}\displaystyle |_{h\rightarrow 0}=v(\tau,\bf{r})
\label{qopid}
\ee
\noindent where in the case when $v(\tau,\bf{r})$  obeys Eq. (\ref{me}) it is called quantum order parameter (QOP) \cite{mukhin}. On the other hand, when $v(\tau,\bf{r})\equiv const (\vec{r})$ it is the common (classical) mean-field order parameter (COP).  Next, differentiating once more Eq. (\ref{dfh}) over $\bar{M}(\tau,\bf{r})$ and using Eq. (\ref{dgm}) one finds \cite{amit}  (we simplify notations by substituting $\tau,\bf{r}\rightarrow x$):

\be
\delta(x,x')=\int dx''\dfrac{\delta^2 F(h)}{\delta h(x)\delta h(x'')}\dfrac{\delta^2 \Gamma(\bar{M})}{\delta \bar{M}(x'')\delta \bar{M}(x')}\Rightarrow \int dx''G^{(2)}(x,x'')\Gamma^{2}(x'',x')|_{h\rightarrow 0}
\label{g2f2}
\ee
\noindent Here we used the fact that in the limit $h\rightarrow 0$  second variational derivative of the generating functional gives two-point Green's function of the order parameter fluctuations according to Eq. (\ref{ZME}):
\be
G^{2}(x,x')=\dfrac{\delta^2 F(h)}{\delta h(x)\delta h(x')}|_{h\rightarrow 0}=\langle( {M}(x)-v(x))(M(x')-v(x'))\rangle
\label{g2}
\ee
\noindent
 Hence, it follows from Eqs. (\ref{g2f2}) and (\ref{g2}), that the second derivative $\dfrac{\delta^2 \Gamma(\bar{M})}{\delta \bar{M}(x'')\delta \bar{M}(x')}=\left[G^{2}(x,x')\right]^{-1}$ is positive semi-definite matrix in Euclidian space and hence, allowing for Eq. (\ref{qopid}), functional $\Gamma(\bar{M})$ is the free energy of the system, as it acquires minimum on the function $\bar{M}(\tau,\bf{r})$ in the order parameter space. 
 Now, for the purpose of merely demonstrating the principle, we assume that $\bar{M}$ is two-dimensional vector and Euclidian action $S_{eff}({M}(\tau,\bf{r}))$ is $O(2)$ symmetric (compare \cite{amit}). Hence, the symmetry-breaking part of the Lagrangian in Eq. (\ref{ZME})  $\propto M'(\tau,\bf{r})h(\tau,\bf{r})$ is invariant under simultaneous rotation by an angle $\theta$ in the symmetry plane  of both $h$ and $M$ expressed in the matrix form:
 
 \be
\hat{T}\vec{h}\cdot\hat{T}\vec{M}=\vec{h}\cdot\vec{M};\; \vec{M}=   \begin{pmatrix} 
      \pi \\
      \sigma \\
   \end{pmatrix};\;  \hat{T}\vec{M} \begin{pmatrix}
         \pi-\theta\sigma \\
         \sigma+\theta\pi \\
      \end{pmatrix}
\label{inv}
\ee
\noindent Hence, the generating functional $F$ is invariant under such simultaneous rotation, because rotating the dummy integrationa variable ${M'}\rightarrow \hat{T}M'$ in Eq. (\ref{ZME}) does not change it:
 \be
\delta F=\theta\int dx\left[\dfrac{\delta F(h)}{\delta h_\sigma}h_\pi-\dfrac{\delta F(h)}{\delta h_\pi}h_\sigma\right]=0;\;
\label{WT1}
\ee
\noindent Substituting relations from Eqs. (\ref{dfh}) and (\ref{dgm}) into Eq. (\ref{WT1})  and allowing for arbitrariness of angle $\theta$, one arrives at the Ward-Takahashi identity \cite{amit}:

\be
\delta \Gamma=\int dx\left[-\dfrac{\delta \Gamma}{\delta \bar{M}_\sigma}\bar{M}_\pi+\dfrac{\delta \Gamma}{\delta \bar{M}_\pi}\bar{M}_\sigma\right]=0
\label{WTG}
\ee
\noindent Now assume that symmetry is broken by magnetic field in $\sigma$ direction:

\be
\vec{h}= \begin{pmatrix}
         0\\
         h_\sigma(x)\\
      \end{pmatrix} \; \rightarrow \vec{M}=   \begin{pmatrix} 
      0\\
      \bar{M}_\sigma(x) \\
   \end{pmatrix}
\label{sigma}
\ee
\noindent Then, variational derivative of the both sides of Eq. (\ref{WTG}) with respect to $M_\pi(x)$ (i.e. transversal mode perpendicular to the magnetic moment, the latter being aligned along $\sigma$ direction) leads to the following equality:
\be
0=h_\pi(x)=\dfrac{\delta \Gamma(\bar{M})}{\delta \bar{M}_\pi(x)}=\int dy\left[-\dfrac{\delta \Gamma}{\delta \bar{M}_\sigma}\delta(x-y)+\dfrac{\delta^2 \Gamma}{\delta \bar{M}_\pi(y)\delta \bar{M}_\pi(x)}\bar{M}_\sigma(y)\right]
\label{preWT}
\ee
\noindent Applying now relation Eq. (\ref{dgm}) to the first term inside the integral in Eq. (\ref{preWT}) one finds:
\be
\int dy\dfrac{\delta^2 \Gamma}{\delta \bar{M}_\pi(y)\delta \bar{M}_\pi(x)}\bar{M}_\sigma(y)=h_\sigma(x);
\label{WTp}
\ee
\noindent that in the limit $h_\sigma(x)\rightarrow 0$ finally leads to the identity:
\be
\int dy\dfrac{\delta^2 \Gamma}{\delta \bar{M}_\pi(y)\delta \bar{M}_\pi(x)}\bar{M}_\pi(y)|_{M^{QOP}(x)}=0.
\label{WT}
\ee
 \noindent in accord with Eq. (\ref{self}). Passing in Eq. (\ref{WT}) to the Fourier transformed functions and allowing for the relation Eq. (\ref{g2f2}) we finally arrive at the generalized Goldstone theorem:  
\be
\bar{M}_\pi(\omega_n)\left[G^{2}(Q,\omega_n)\right]^{-1}=0;\;\omega_n=2\pi nT 
\label{preG}
\ee
\noindent where bosonic frequencies arise from the periodicity of the QOP function defined by Eq. (\ref{SHM}). Now we immediately see from Eqs. (\ref{me})  and  (\ref{SHM}) that "Goldstone modes" of the QOP do not necessarily possess pole at zero frequency, since $\bar{M}^{QOP}_\pi(\omega_n=0)=0$ and relation Eq. (\ref{preG}) is then fulfilled already at finite $G^{2}(Q,\omega_n=0)$.  On the other hand, for a finite Matsubara frequencies $\omega_n=2\pi nT$  the Green's function has poles according to Eq. (\ref{preG})!  The QOP period $T_{QOP}$ along the Matsubara axis remains finite: $T_{QOP}\rightarrow 1/nT=const$ in the limit $T\rightarrow 0$, see \cite{mukhin} and next Section. Hence, there exists a characteristic  \textit{finite} Matsubara frequency $\omega_0=2\pi T_{QOP}$ , for which there is a pole: $\left[G^{2}(Q,\omega_0)\right]^{-1}=0$.
We observe here, that while analytically continued to the real axis of frequencies $\omega$ this may give rise to a finite-frequency resonance in the transverse spin-modes, which might be associated with the "neutron resonance" peak measured \cite{hg,hg1,hg2,hg3} at finite frequency in the magnetic (bosonic) excitations spectrum in lightly hole-doped copper oxides at antiferromagnetic point $\vec{Q}=\{\pi,\pi\}$ .

\section*{3. Quantum ordered photonic state of coupled Josephson array}
\label{sec: JJ-QOP}
It was predicted recently \cite{mf} that QOP state may be realized in the system of  Josephson junctions (JJ) in a resonant cavity considered as long-range coupled two-level systems (TLS). The effective Euclidian action of TLS coupled to the photon field could be derived in analogy with Eqs. (\ref{ZM}), (\ref{SF}) as one notices that array of TLS systems can be mapped on the gas of Fermi-oscillators, while electromagnetic field in the cavity is a natural realization of the Hubbard-Stratonovich field that transfers interaction between different TLS:
 
\be
 S_{eff}[P(\tau)]=\int_0^{\beta}d\tau \frac{1}{2m}\left[P^2+\frac{1}{\omega_0^2} \dot P^2\right]-k_BT\sum_i \ln \left[\cosh{\alpha_i\{P\}}\right]
\label{tls}
\ee
\noindent The Floquet eigenvalues $\alpha_i(P)$ are now determined from the following equation \cite{mf}:
\be
(\partial_\tau+\hat H^{i}_{P})\psi_i=0;\;   \psi_i(\tau+{1}/{k_BT})=e^{-\alpha_i}{\psi}_i(\tau)
\label{Equation-Floquet}
\ee
\noindent and the Hamiltonian $H^{i}_{P}$ is :

\begin{equation}
\hat H^{i}_{P}=\left(
    \begin{array}{cc}
      \epsilon_i+\tilde \eta_i P(\tau) & \Delta_i\\
      \Delta_i & -(\epsilon_i+\tilde \eta_iP(\tau))
    \end{array}
  \right)~,
 \label{Hamiltonian-Floquet}
\end{equation}
where $\tilde \eta_i$ is the parameter determining the interaction between cavity modes and two-level systems , and $\Delta_i$ and $\epsilon_i$ are the matrix element of tunneling and the energy difference between two potential wells, accordingly (see \cite{mf} for details).  Then, the  role of  the QOP periodic in Matsubara's time is played by the photons momentum  $P_{0}(\tau+{1}/{k_BT})=P_{0}(\tau)$, that  minimizes the effective action and is a solution of the equation analogous to Eq. (\ref{self}) :
\begin{equation} 
\frac{P_0}{m}+\frac{1}{m\omega_0^2}{\partial_\tau}^2 P_0=\sum_i \frac{\partial \alpha_i}{\partial P} th \left(\frac{\alpha_i\{P_0\}}{2}\right)~.
\label{Min-effectivaction}
\end{equation}
\noindent here index $i$ enumerates the TLS in the array. Here the Hamiltonian of a photon field has the following second quantized form:
\be 
H_{ph}=\hbar  \omega_0(\hat b^+ \hat b+1/2)~, \omega_0=ck_n.
\label{Hamiltonian-Photon-2}
\ee
\noindent which assumes the following expressions for the conjugated operators of the electromagnetic field expressed via creation and annihilation operators of resonant photons in a cavity:
\be
\hat Q(t)=\sqrt{\frac{\hbar}{ck_nL_0 \ell}}(\hat b+\hat b^+)
\label{Q}
\ee
\noindent and
\be
\hat P(t)=-i\sqrt{\frac{\hbar ck_nL_0 \ell }{4}}(\hat b-\hat b^+)~.
\label{P}
\ee
\noindent Here $L_0$ and $m=L_0l/2$ designate inductance density and the total inductance of the TLS array (of length $l$) in the cavity; and $c$ and $k_n$ are velocity and wave-vector of light respectively. It is also assumed below for simplicity that photon's wave-length $1/k_n$ is much greater than the cavity size, so that all the TLS are coupled to the space-independent photon field. Now, it is possible to solve analytically the whole problem expressed in Eqs. (\ref{tls})-(\ref{Min-effectivaction}) in the simplistic, but still nontrivial case:
$\epsilon_i=0$, $\Delta_i\equiv\Delta$ and $\tilde \eta_i=\tilde \eta $. Thus, we consider a phase transition into a quantum ordered state representing a quantum interference between the two semi-classical states  $\pm P_0$ of the photon field, which are inversion symmetry related. Each of the $\pm P_0$ states separately describes a particular coherent state of the photon field. The quantum ordered state is then described by the Matsubara's time-dependent amplitude of the semi-classical photon field $P_0(\tau)$, playing the role of QOP \cite{mukhin}
(quantum order parameter), which is the periodic in imaginary time solution of Eq. (\ref{Min-effectivaction}). We consider analytic solution first found in \cite{mukhin,mukhin1}:
\begin{equation} \label{Selfcons-tau-P0}
\tilde \eta P_0(\tau)= 4nk_{B}KTk_1 sn\left( {4nk_{B}KT\tau ;k_1 } \right),\,K = K(k_1 )
\end{equation}
where $sn(\tau,k_1)$ is the Jacobi snoidal elliptic function, periodic in $\tau$ with period $1/(nT)$, $n=1,2,...$, (the Boltzman constant is taken for unity), $K(k_1 )$ is elliptic integral of the first kind, positive integer $n$ and  real parameter $0< k_1 <1$ are found by minimizing the Euclidian action $S_{eff}$ given in Eq. (\ref{tls}). For this ansatz the Floquet index is found in analytical form in \cite{mukhin,mukhin1}:
\be
{\alpha}=2\varepsilon\left(\dfrac{1-{{k}}^2+\varepsilon^2}{1+\varepsilon^2}\right)^{1/2}
n\Pi\left(\dfrac{{{k}}^2}{1+\varepsilon^2},{{k}}\right);\;\varepsilon=\dfrac{\Delta}{2nK(k)T}.
\label{flon}
\ee
\noindent Substituting Eqs. (\ref{Selfcons-tau-P0}) into Eq. (\ref{tls}) and assuming $\omega_0\gg nKT$ one can integrate explicitly in Eq. (\ref{tls}) and also substitute summation over TLS index $i$ with multiplication by factor $N$ - the number of TLS in the array:
 \be
S_{eff}[P(\tau)]=\dfrac{(2nK(k)T)^2}{4UT}\Theta(k)-N\ln\cosh(\alpha/2)&&
\label{gl}\\
0\leq\Theta(k)\equiv 1+k'^2-2\dfrac{E(k)}{K(k)}\leq 1; k'^2=1-k^2&&
\label{s}
\ee
\noindent where the left and right limits of $\Theta(k)$ are achieved at $k=0$ and $k=1$ respectively, and $l$ is the length of the TLS system, $E(k)$ is complete elliptic integral of the second kind. Parameter $U=m{\tilde \eta}^2$ is effective local coupling of a TLS to the photon field. As usual, $T$ designates the temperature of the system.  In the strongly nonlinear limit $k'\rightarrow 0$, that arises in the case $UN\gg \Delta$, the Floquet index can be expressed in a simple asymptotic form:
\be
\alpha\approx 2n\ln{\left(\dfrac{\varepsilon}{k'}+\sqrt{1+\dfrac{\varepsilon^2}{k'^2}}\right)}
\label{alpha}
\ee 
\noindent Substituting (\ref{alpha}) into (\ref{gl}) one finds the following expression for the effective Euclidian action $S_{eff}$ as a function of the new parameters to be determined by minimization of $S_{eff}$:
\be
S_{eff}[k',n]=\dfrac{(XK)^2}{4UT}\left( 1-\dfrac{2}{K}\right)-Nn\left(\ln\left\{\dfrac{2\Delta}{XK}\right\}+K\right);\; X=2nT;\; K\equiv K(k').
\label{eua}
\ee
\noindent A direct minimization of $S_{eff}$ in (\ref{eua}) with respect to variables $X,K$ leads to the following final result:
\be
n_{QOP}=\dfrac{\Delta}{4Te};\;  k'_{QOP}=4\exp\left\{-\dfrac{2UNe}{\Delta}\right\};\; S_{eff}^{QOP}\equiv S_{eff}(k'_{QOP},n_{QOP})=-\dfrac{UN^2}{4T}
\label{qopsol}
\ee 
\noindent This result is remarkable, since it demonstrates that in the QOP state the period of the "Euclidian crystal" along the Matsubara's exis , $1/(2nT)$, remains finite as the temperature goes to zero, hence, "Euclidian crystal"  survives at zero temperature.  It is important to compare QOP solution in Eq. (\ref{qopsol}) with the classical mean-field solution $P_{COP}(\tau)\equiv P_0=const$.
In this case Floquet index and effective action equal:
\be
\alpha_{COP}={\dfrac{\sqrt{\Delta^2+P_0^2}}{T}}; \;\;S_{eff}^{COP}=\dfrac{P_0^2}{4UT}-N\ln{ch{\dfrac{\sqrt{\Delta^2+P_0^2}}{2T}}}
\label{cla}
\ee
\noindent Minimization of the above expression for $S_{eff}^{COP}$ leads to the following results:
\be
P_{COP}\approx UN;\;\;S_{eff}^{COP}\approx -\dfrac{UN^2}{4T}.
\label{clasol}
\ee
\noindent Hence, comparing Eqs. (\ref{qopsol}) and (ref{clasol}) we conclude, that QOP solution manifests  another local minimum of the free energy of the photon field, besides the  COP state's minimum. A more detailed calculation is necessary to compare these two energies with the higher accuracy. Finally, it is interesting to study how COP and QOP states are destroyed by the quantum disordering. The latter could be introduced in analogy with the anti-nesting parameter $t$ in \cite{mukhin}, thus leading to the following modified Euclidian action of the photon field:
\be
S_{eff}[P(\tau)]=\dfrac{(2nK(k)T)^2}{4UT}\Theta(k)-\frac{N}{2}\left[\ln\cosh(\alpha/2+\dfrac{t}{2T})+\ln\cosh(\alpha/2-\dfrac{t}{2T})\right]
\label{glt}
\ee
\noindent Minimization of $S_{eff}[P(\tau)]$ in Eq. (\ref{glt}) leads to the conclusion that $P_{QOP}$ vanishes abruptly (i.e. via first order quantum transition) at the critical value of $t=t_c^{QOP}$:
\be
t_c^{QOP}={UN}.
\label{tcqop}
\ee
\noindent Simultaneously, the anti-nesting parameter $t$ modifies Euclidian action of the photon field for the COP case as:
\be
S_{eff}[P_0]=\dfrac{P_0^2}{4UT}\Theta(k)-\frac{N}{2}\left[\ln\cosh(\dfrac{\sqrt{\Delta^2+P_0^2}+t}{2T})+\ln\cosh(\dfrac{\sqrt{\Delta^2+P_0^2}-t}{2T})\right]
\label{clt}
\ee
\noindent Minimization of $S_{eff}[P_0]$ in Eq. (\ref{clt}) leads to the conclusion that $P_{COP}$ vanishes abruptly (i.e. via first order quantum transition) at the critical value of $t=t_c^{COP}$:
\be
t_c^{QOP}={UN}.
\label{tcqop}
\ee
\noindent Hence, the QOP state exists in the same domain as the COP phase in the $P-t$ diagram. The fine structure of the domains boundary has to be calculated.

\section*{4. Conclusions}
\label{sec: fin}
It is demonstrated that quantum ordering in the system with global continuous symmetry, contrary to the case of classical ordering characterized with emergence of a  mean-field order parameter, \textit{does not cause necessarily} appearance of massless Goldstone bosons (i.e. gapless Goldstone modes).  It is demonstrated that in the quantum ordered state the  Goldstone theorem does apply to the modes at nonzero Matsubara frequencies, i.e. after Wick rotation to the real axis the Goldstone modes may possess a finite gap. This result leads to important consequences when considered in combination with another peculiar properties of the QOP state found recently \cite{mukhin}: {\it{QOP does not scatter anything in Minkowski world}}! Hence,  the present theory, we believe, may naturally indicate that two effects:  "hidden order" revealed in ARPES spectra of fermi-excitations  \cite{pg1,pg2} and "neutron resonance" feature \cite{hg,hg1,hg2,hg3} at finite frequency in the magnetic (bosonic) excitations spectrum in lightly hole-doped copper oxides, could be emerging fingerprints of the QOP state. A peculiar manifestation of the QOP self-consistent solution is also predicted for a photon field coupled with the array of Josephson junctions  in the superconducting resonator cavity. It is demonstrated that it constitutes a competing (meta)stable state of the system, that, for a particular quantum disordering model, and to zero order accuracy in small parameter $\Delta/UN$ exists inside the same stability domain of the classical photonic condensate delimited by the critical value of the quantum disordering strength.

\section*{Acknowledgement}
The author acknowledges useful discussions with Prof. Jan Zaanen and Dr. Michail Fistul, as well as partial support of this work by the Russian Ministry of science and education grant No. 14A18.21.1936,  MISIS grant 3400022 , and RFFI grant 12-02-01018.

\end {document}